\documentclass[preprint,12pt]{elsarticle}

\usepackage{lineno}

\usepackage[dvipdfmx]{color}

\usepackage{graphicx}

\usepackage{amssymb}

\journal{Physics Letters B}

\begin{document}

\begin{frontmatter}



\title{Observation of gamma ray bursts at ground level under the thunderclouds}


\author[minowa]{Y.~Kuroda}
\author[KEK]{S.~Oguri}
\author[minowa]{Y.~Kato}
\author[minowa]{R.~Nakata}
\author[icepp]{Y.~Inoue}
\author[JAEA]{C.~Ito}
\author[minowa]{M.~Minowa\corref{cor1}}
\ead{minowa@phys.s.u-tokyo.ac.jp}

\cortext[cor1]{Corresponding Author}
\address[minowa]{Department of Physics, School of Science, The University of Tokyo,
  7-3-1, Hongo, Bunkyo-ku, Tokyo 113-0033, Japan}
\address[KEK]{Institute of Particle and Nuclear Studies,
  High Energy Accelerator Research Organization (KEK), 1-1, Oho, Tsukuba, Ibaraki 305-0801, Japan}
\address[icepp]{International Center for Elementary Particle Physics, The University of Tokyo,
  7-3-1, Hongo, Bunkyo-ku, Tokyo 113-0033, Japan}
\address[JAEA]{Oarai Research and Development Center, Japan Atomic Energy Agency,
  4002, Naritacho, Oarai-machi, Higashiibaraki-gun, Ibaraki 311-1393, Japan}

\begin{abstract}
We observed three $\gamma$-ray bursts related to thunderclouds in winter
using the prototype of anti-neutrino detector PANDA made of 360-kg plastic scintillator
deployed at Ohi Power Station at the coastal area of the Japan Sea.
The maximum rate of the events which
deposited the energy higher than 3\,MeV was
$(5.5 \pm 0.1) \times 10^2 {\rm /s}$.

Monte Carlo simulation showed that electrons with approximately monochromatic energy
falling downwards from altitudes of order 100\,m roughly produced the observed total energy spectra of the bursts.
It is supposed that
secondary cosmic-ray electrons, which act as seed,
were accelerated in electric field of thunderclouds and
multiplied by relativistic runaway electron avalanche.
We actually found that the $\gamma$-rays of the bursts entered into the 
detector from the direction close to the zenith.
The direction stayed constant during the burst within the detector resolution.

In addition, 
taking advantage of the delayed coincidence detection
of the detector, we found  neutron events in one of the bursts 
at the maximum rate of $\sim 14\pm5\,{\rm /s}$.
\end{abstract}


\end{frontmatter}



\section{Introduction}

In the early 1920's, C.T.R. Wilson suggested that 
strong electric fields in thunderclouds might 
accelerate free electrons present in the atmosphere
to high energies\cite{Wilson1924}. 
Since then, radiation associated with thunderstorms 
attracted the interest 
as natural particle-acceleration process
and many experiments have been attempted 
to detect these radiations in various environments.

For instance, bursts of $\gamma$-rays were observed on orbiting satellites with energy up to tens of MeV
and with duration of less than 1 ms.
They are called Terrestrial Gamma-ray Flashes(TGF's)\cite{TGF}. 

Recently, Dwyer {\it et al.}~\cite{Dwyer2015} reported unexpected observation of 
positron bursts, lasting about 0.2\, s,
by an airborne detector when the aeroplane flew into a thundercloud.

On the other hand, $\gamma$-ray flux enhancements of longer duration of order 100 s
were reported in limited environments like high mountains~\cite{Suszcynsky1996, Brunetti2000, Chubenko2000,
Chubenko2003, Alexeenko2002, Torii2009, Tsuchiya2009, Tsuchiya2012, Chilingarian2010, Chilingarian2012} and  sea level locations in the coastal area of the Japan Sea.
They are also called Thunderstorm Ground Enhancements(TGE's)~\cite{Chilingarian2012}.
Japanese groups found that 
radiation monitoring posts or dedicated scintillation counters in and near nuclear power plants signaled 
an increase of $\gamma$-ray dose 
which seemed to originate from low altitude winter thunderclouds\cite{Yamazaki2002,Torii2002,Enoto2007,Tsuchiya2007,Tsuchiya2011,Tsuchiya2013}.
Especially, Torii {\it et al.}~\cite{Torii2011} found that area of $\gamma$-ray flux enhancements
was moving as the associated thundercloud passed across the observation site.

Gurevich {\it et al.}~\cite{Gurevich1992,Gurevich2005} developed
the runaway electron model to explain the electron acceleration 
in the electric field of the thunderclouds.
The stopping power of air for electrons decreases with increasing electron energy and goes up again by relativistic effects.
Therefore, electric field in the thundercloud may accelerate electrons if the the electric force is larger than the minimum stopping power and
the electron energy is in the region where the electric force exceeds the stopping power.  
Such electrons are called runaway electrons. 
By generating knock-on electrons successively, the runaway electrons can cause an avalanche multiplication process called
relativistic runaway electron avalanche (RREA).

Numerical simulations\cite{Torii2004a, Babich2010a, DwyerPR2014} 
with models of thundercloud electric field
suggested that the avalanche can be produced 
continuously 
if energetic seed electrons are provided, for example, by cosmic ray secondaries.
A significant flux of relativistic runaway electrons in the lower parts of thunderclouds
is capable of producing intensive bremsstrahlung which can reach the
Earth's surface or the mountain top to account for the observed flux enhancement.

Recently, the neutron bursts associated with thunderstorms were
also observed in various experiments~\cite{Shyam1999,Chilingarian2010,Martin2010,Gurevich2012,Starodubtsev2012}.
The generation of neutrons is most probably by photoproduction by $\gamma$-rays with
air nuclei as the detected $\gamma$ ray spectrum extends 
above the photonuclear reaction threshold for nitrogen ($\sim 10.5 {\rm MeV}$)~\cite{Babich2010}.
It may have a significant effect on 
${}^{14}{\rm C}$ dating~\cite{Libby1973,Fleischer1974} through the neutron capture reaction
${}^{14}{\rm N}(n,p) {}^{14}{\rm C}$.

Our research group have developed prototypes of 
a reactor neutrino detector ``PANDA'',
which stands for Plastic Anti-Neutrino Detector Array~\cite{LesserPANDA,PANDA36}.
We have originally targeted PANDA at presenting
the feasibility of reactor monitoring using neutrinos
with a tonne-size detector.
$\gamma$-rays and neutrons can also be detected by PANDA by Compton scattering
and the delayed coincidence of proton recoil and neutron captures.
We installed the PANDA detector 
outside of the reactor building of Ohi power station, which stands near the Japan Sea in Fukui,
and tried to watch the reactor operating status
via detecting and analyzing the anti-electron neutrinos produced in the reactor core.

We accidentally found that
there were intensive increases of $\gamma$-ray flux
correlated with the winter thunder-storm activity
during the measurement.
In this paper, we report the
investigated properties of these burst events
taking advantages of the unprecedented features of the detector
including high statistics, good energy response,
direction sensitivity and neutron identification.

\section{Experimental setup}

%
%

Our prototype detector ``PANDA36'' consists of thirty-six (six by six) stacked modules~\cite{LesserPANDA,PANDA36}.

%
%

\begin{figure}[tbp]
  \begin{center}
    \includegraphics[width=\textwidth]{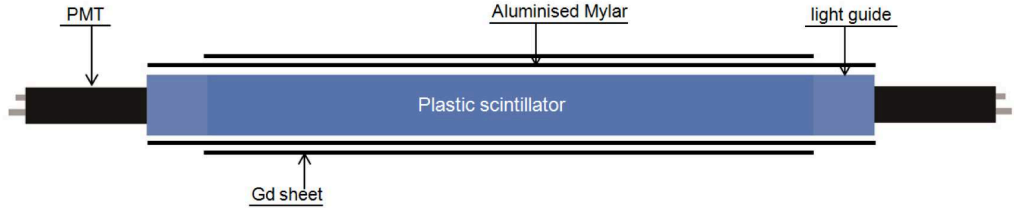}
    \caption{Structure of the PANDA module}
    \label{fig:PANDA_module}
  \end{center}
\end{figure}

The module was made of 
a plastic scintillator bar($10 {\rm cm} \times 10 {\rm cm} \times 100 {\rm cm}$) with effective mass of about 10kg wrapped 
with aluminized Mylar films and gadolinium (Gd) coated Mylar films (4.9 mg of Gd per ${\rm cm}^2$).
Each bar was connected to acrylic light guides and photomultipliers on 
both ends (Figure~\ref{fig:PANDA_module}).

The light intensity ratio seen by each PMT pair allows one 
to estimate the position of the hit along the module~\cite{LesserPANDA}.
Using the position of the hit and the charge outputs from each PMT,
one can estimate the energy deposit of the hit.
The position and energy resolutions
were 16\,cm and 300\,keV for
4\,MeV hit on the center, respectively.

Each PMT signal was divided into two:
about $15\%$ of the original charge was sent
to CAEN V792 multi-event Charge-to-Digital-Converters (QDCs)
and the other $85\%$ was passed to
CAEN V895 leading edge discriminators.

The discriminator outputs were sent to CAEN V1495 general purpose VME board,
which has customizable FPGA unit (Altera Cyclone EP1C20).
The logic counted the number of pairs of fired PMTs seeing the same scintillator.
Whenever the number of the pairs was greater than
or equal to two, the logic generated the gate pulses of 400\,ns duration
for the QDCs.

The timing of the gate pulses and busy signals from the QDCs were
recorded by the same FPGA.
We used these time stamps to select neutrino events 
by delayed coincidence method offline.

The PANDA36 detector was loaded on and transported by a 2-tonne dry van.
The detector was deployed beside the Unit 2 of Ohi Power Station 
($35^{\circ}32'32''{\rm N}, 135^{\circ}39'14''{\rm E}$
and about 10~m above the sea level)
of Kansai Electric Power Co., Inc on November 18th, 2011.
We continued the measurement for 62 days.

Energy calibrations were carried out before the deployment using the Compton edge of  $^{60}$Co $\gamma$-rays.
Time drifting of gains of each PMT and QDC was corrected using the peak of through-going cosmic muons in the spectrum of the events.

\section{Event-rate increase}

In the data acquired 
by the PANDA36 detector through the neutrino detection experiment,
we found unexpected increases of event rate.
The trigger rate got twice or higher for a few minutes
for the events with total energy deposit larger than 3\,MeV
independently of the reactor operation.

\begin{figure}
    \mbox{\includegraphics[width=7cm]{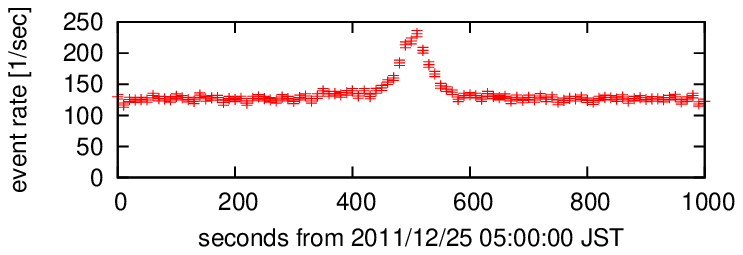}}
    \mbox{\includegraphics[width=7cm]{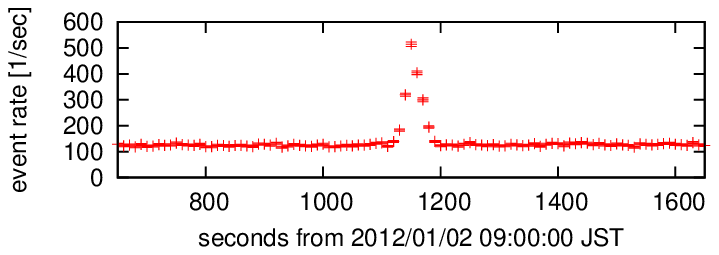}}
    \mbox{\includegraphics[width=7cm]{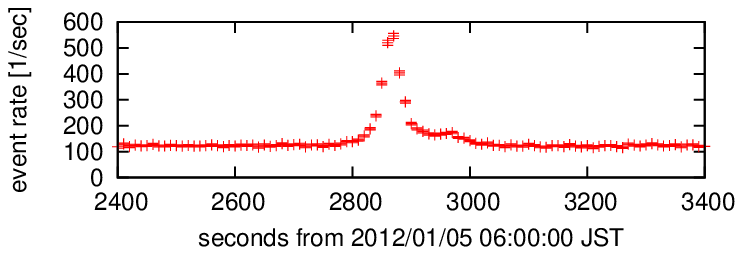}}
    \caption{Temporal variation of the event rates of three bursts}
    \label{fig:eventrates}
\end{figure}

Temporal variation of the event rate are shown in 
Figure~\ref{fig:eventrates}.
Burst duration is defined as an interval whose event rates are
5$\sigma$ greater than average event rate.
The date and time, duration and peak event rate are summarized in Table~\ref{tab:features}.
We hereafter call each burst by the name defined in the table.

We searched for the association between thunderclouds and the radiation bursts
via ``Kaminari Nowcast'' data provided by Japan Meteorological Agency.
Kaminari Nowcast provides the thunder activity information
analyzed from lightning discharges detected by Lightning Detection Network System (LIDEN) and
radar observations for every $\sim1\,{\rm km}$ grid~\cite{JapanMeteorologicalAgency}.

Kaminari Nowcast data show that there were more than one grid 
which have level 2 thunder activity of five levels around the experimental site for all three bursts
at almost the same time that the radiation bursts have been observed.

Conversly, we found 22 time-consecutive data sets of level 2 or higher 
in 20 $\times$ 20 grids around the detector.
Nevertheless, radiation bursts are observed only three times.
It is not strange since the $\gamma$-ray emitting region is conceivably much smaller than the above area.

\section{The energy and the height of the source electrons}\label{sec:source}

We performed Monte Carlo simulations to estimate
the energy and the height of the source electrons using Geant4 toolkit~\cite{Agostinelli2003},
assuming the $\gamma$-rays are caused by high energy electrons emitted downwards from
thunderclouds.

At first, 
we simulated the numbers and spectra of bremsstrahlung photons at the ground level
initiated by mono-energetic electrons projected vertically downward from the sky.
The height and energy of the source electrons were chosen to be combinations of
100\,m, 500\,m, 900\,m, 1300\,m,
and 17\,MeV, 23\,MeV, 29\,MeV, 35\,MeV.
As a generic spectral shape of the source electrons,
we estimated it as a sum of these four components.
Lower-energy electron components were ignored 
for the efficiency of analysis.
It does not necessarily exclude the theoretically expected exponential electron energy distribution~\cite{Dwyer2011}.  
Lower-energy components of electrons, if exist, are less efficient in producing bremsstrahlung $\gamma$-rays on the ground, and therefore flux of those lower-energy electrons could only be determined with poor accuracy.

Then we calculated the detector response to those $\gamma$-rays.
Next, we calculated the weighted sum of the spectra of the detector responses
to the simulated bremsstrahlung $\gamma$-rays from four energies of electrons
projected from each height. 

\begin{figure}[htbp]
    \mbox{\includegraphics[width=7cm]
    {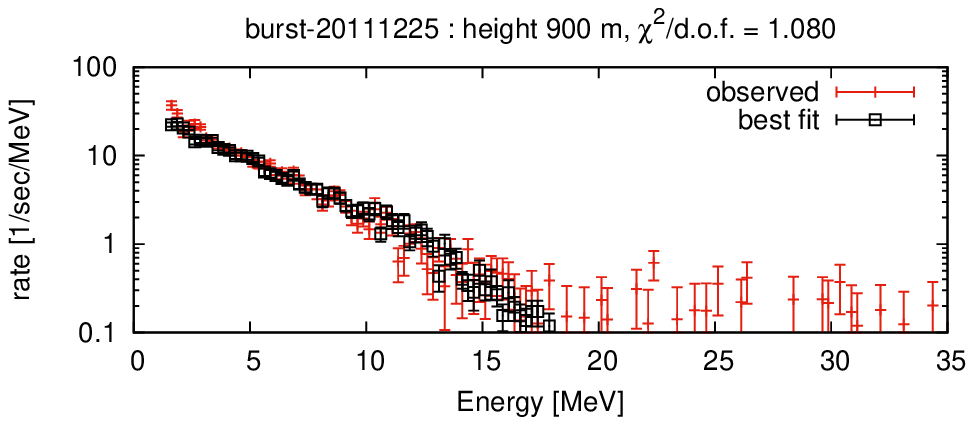}}
    \mbox{\includegraphics[width=7cm]
    {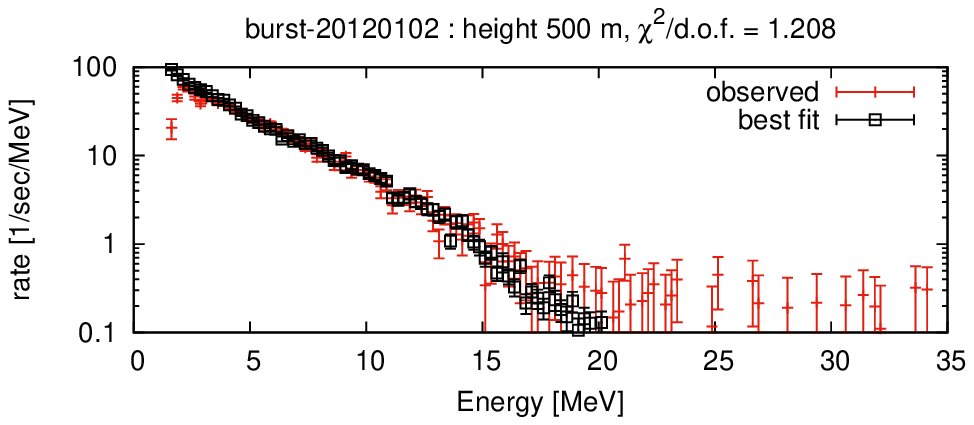}}
    \mbox{\includegraphics[width=7cm]
    {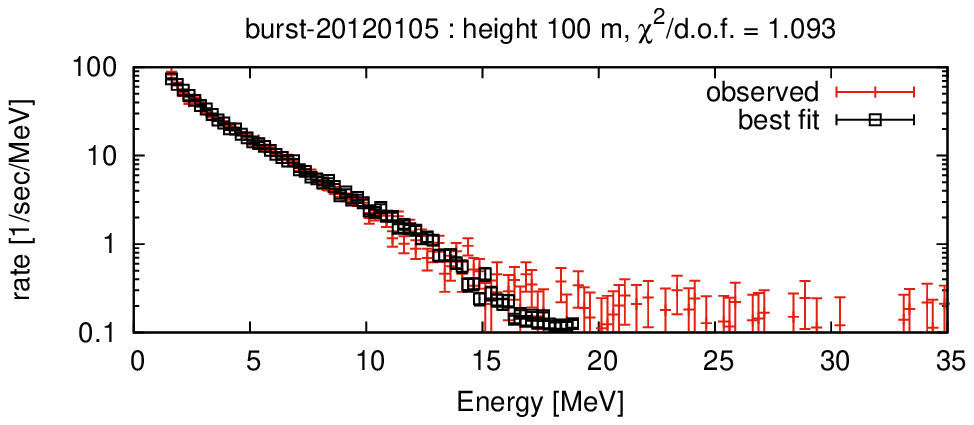}}
    \caption{Simulated spectra which match the observed burst spectra best.  The error bars shown are statistical.}
    \label{fig:bremss_spectrum}
\end{figure}

\begin{figure}[htbp]
    \mbox{\includegraphics[width=7cm]
    {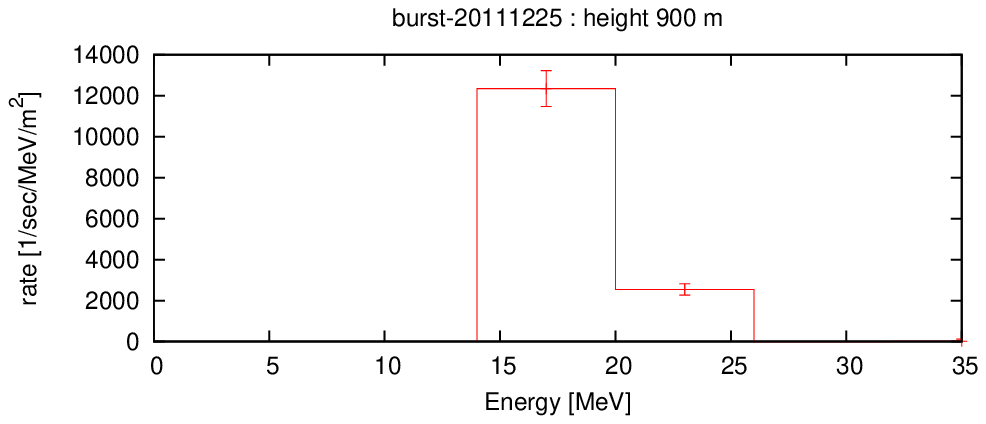}}
    \mbox{\includegraphics[width=7cm]
    {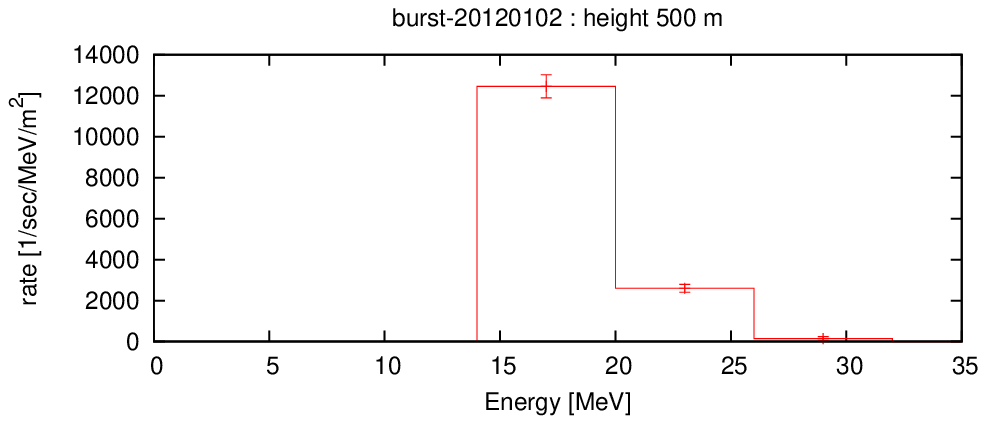}}
    \mbox{\includegraphics[width=7cm]
    {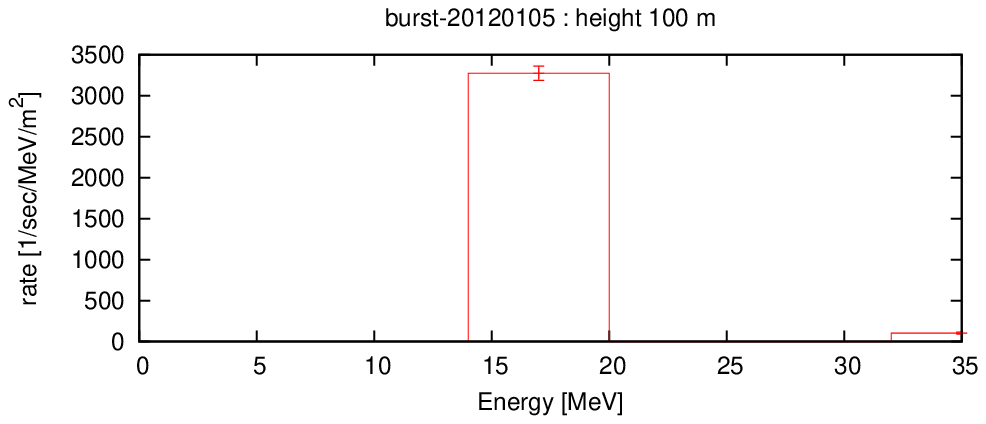}}
    \caption{Estimated source electron spectra at the projection height with statistical errors of three bursts}
    \label{fig:electron_spectrum}
\end{figure}

The summed spectra were compared with the observed spectra of the bursts
after background subtraction.
The background data were taken in two ten-minute intervals
starting
thirteen minutes before and three minutes after the burst periods.
It is seen that the observed spectra extend up to 15 MeV or more.
We fitted the weights to minimize the $\chi^2$
for each projection height.
The analyses were done for all the three bursts and the 
fit results with the heights and the weights
which minimize $\chi^2$ are shown in 
Figure\,\ref{fig:bremss_spectrum}.
The source electron spectra with the weights at the projection heights are shown in 
Figure\,\ref{fig:electron_spectrum}.

\begin{figure}[htbp]
  \begin{center}
    \includegraphics[width=7cm]
    {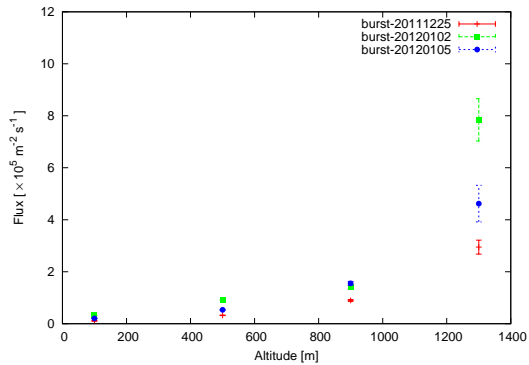}
    \caption{Estimated peak flux of the source electrons as a function of the assumed height}
    \label{fig:flux_vs_height}
  \end{center}
\end{figure}

They indicate that 17 MeV electrons, as a somewhat arbitrary choice, produce a spectrum qualitatively similar to the data.
Peak flux of the source electrons of each burst is 
$(1.9\pm 0.1)\times 10^{5}{\rm m}^{-2}{\rm s}^{-1}$, 
$(2.0\pm 0.1)\times 10^{5}{\rm m}^{-2}{\rm s}^{-1}$ and 
$(8.8\pm 0.5)\times 10^{4}{\rm m}^{-2}{\rm s}^{-1}$,
for burst-20111225, burst-20120102 and burst-20120105, respectively.  
It should be noted that 
the estimated source electron fluxes are 
the lower limits
since we ignored the electron components lower than 17MeV.
Estimated peak flux of the source electrons is also plotted as a function of the assumed height in Figure\,\ref{fig:flux_vs_height} to see the dependence of the flux on height.

\section{Arrival direction of $\gamma$-rays}

The arrival direction of a $\gamma$-ray can be investigated,
if the $\gamma$-ray is Compton-scattered by an electron in a plastic scintillator module
and then the scattered $\gamma$-ray deposits most of its energy in another module.
The segmented structure of PANDA36 detector
allows one to estimate the energy deposit and the position of the deposit
along the module.

The arrival direction lies along the cone called the Compton cone with its axis being the line connecting two interaction points and its half opening angle $\alpha$ which satisfies the relation,
\begin{equation} \label{eq:compton_scattering}
  E'_{\gamma}=\frac{E_{\gamma}}{1 + \frac{E_{\gamma}}{m_ec^2}(1-\cos\alpha)},
\end{equation}
where $E_{\gamma}$ is the energy of the incident photon and
$E'_{\gamma}$ is the energy of the scattered photon.

For the analysis, 
we chose the events with the total energy deposit in the range $5\,{\rm MeV} \le E_{\rm total} \le 12\,{\rm MeV}$
to remove the ordinary environmental $\gamma$-rays 
and to take those $\gamma$-rays which deposited enough energy on each module 
to ensure sufficient position resolution.
We assumed $E_{\rm total}$
to be the incident $\gamma$-ray energy, $E_{\gamma}$.
We also assumed that  $E_{\gamma} - E'_{\gamma}$ and $E'_{\gamma}$ correspond to
 $E_{\rm 1st}$ and $E_{\rm 2nd}$, the highest and the second highest energy deposit 
 of all the modules, or vice versa.
We simply chose the combination  which corresponds to the gamma ray coming from the upper hemisphere.

The first assumption that $E_{\rm total}=E_\gamma$ is often not true.
To reduce the effect of error in $E_{\rm total}$,
we introduced the selection criterion,
\begin{equation}\label{eq:compton_cut}
  \frac{E_{\rm 1st}}{2} \le E_{\rm 2nd} .
\end{equation}
We can cut the energy region near the Compton edge by this selection ,
where the effect of $E_{\rm total}$ error to $\cos\alpha$ is large.
The selection can effectively cut the event with two Compton scatterings in one module, too.

In addition, the positions of $E_{\rm 1st}$ and $E_{\rm 2nd}$ were required 
to be interleaved with more than two modules along the stacking direction.
This selection reduces the effect of the periodical structure of the detector
which may discretize the result of the scattering angle calculation.

After above preparations,
the Compton cone and the corresponding circle on the unit sphere centered at the detector center
was calculated for each event which satisfied the selection criteria.
Then each event is counted in the predefined grids in $(\cos\theta,\phi)$ space
with a weight of the fraction of the circumference which overlaps the grid.
The polar angle $\theta$ and azimuth $\phi$ are defined with respect to the axis along the length of the modules
so that the zenith is (0,$\pi$/2).
Then we normalized the weighted number of events in each grid by the live time of the burst so that
it represents the arrival rate of the selected event from the direction.

\begin{figure}[tbp]
     \mbox{\includegraphics[width=5cm]{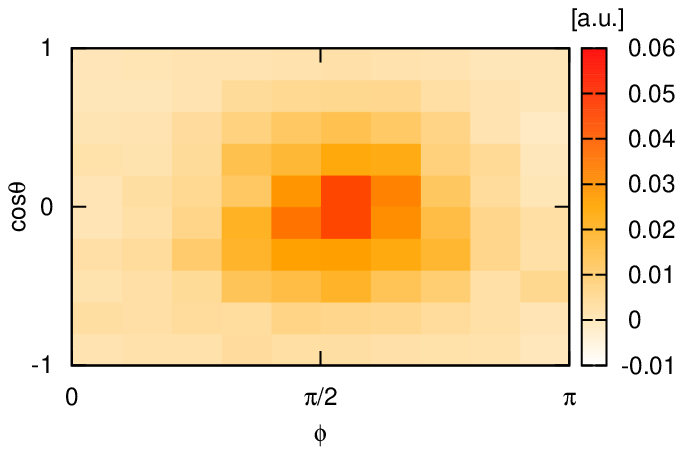}}
     \mbox{\includegraphics[width=4cm]{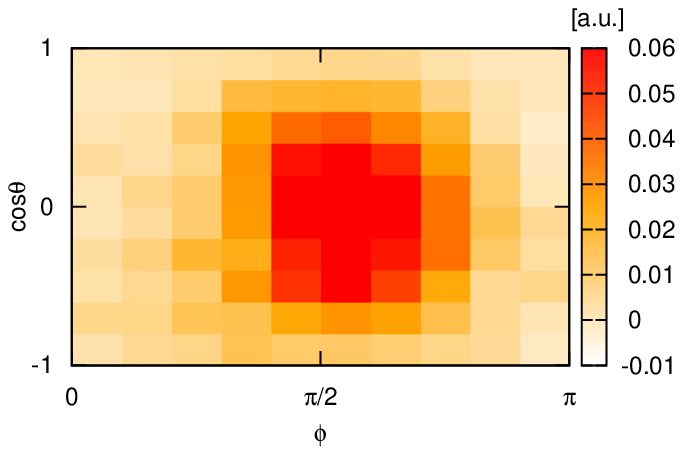}}
     \mbox{\includegraphics[width=4cm]{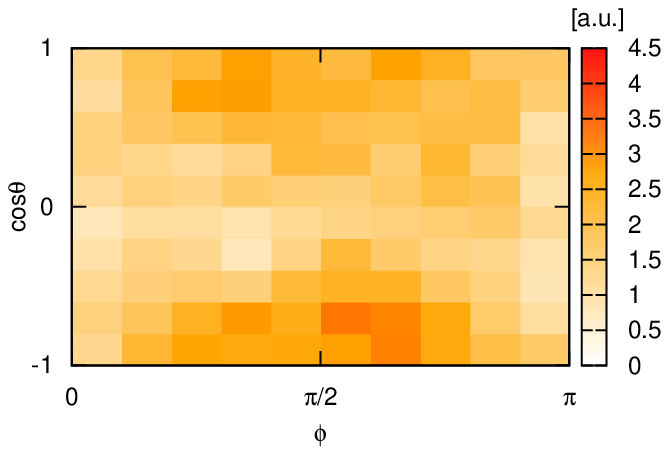}}
   \caption{Arrival direction of burst-20120105 with detector response correction(left), 
   		without correction(middle) and detector response for $\gamma$-rays 
		isotropically incident on the detector(right)}
    \label{fig:direction}
\end{figure}

Monte Carlo simulations were made for $\gamma$-rays isotropically incident on the detector
and then the same arrival-direction analysis was made as for the data.
We made the following calculation to get rid of a possible bias due to the detector response,
\begin{equation}\label{eq:arrival_direction}
  M_i = (M_{i,{\rm burst}} - M_{i,{\rm bg}}) / (M_{i,{\rm sim}}).
\end{equation}
Here $i$ represents the grid number,
$M_{i,{\rm burst}}$, $M_{i,{\rm bg}}$ are the arrival rate of each grid
in the burst period and the background period, respectively.
$M_{i,{\rm sim}}$ is the arrival rate calculated by the Monte Carlo simulation,
which is normalized so as to get the average value of all the grids to be 1.
Maps of $M_i$, $M_{i,{\rm burst}} - M_{i,{\rm bg}}$ and $M_{i,{\rm sim}}$ for 
burst-20120105 are shown in Figure\,\ref{fig:direction} as typical examples.

We found the arrival direction was close to the zenith as shown in 
Figure\,\ref{fig:direction} and
stayed constant during the periods of all three bursts within the rate of 
$\frac{d\cos\theta}{d t} \leq (0.2 \pm 0.5)/30{\rm s}$ and 
$\frac{d\phi}{d t} \leq (0.4 \pm 0.5){\rm rad}/30{\rm s}$.

\section{Neutron event-rate increase}

PANDA36 detector is capable of detecting fast neutrons by delayed coincidence method.
A neutron entering the detector interacts with a proton in the plastic scintillator.
A neutron transfers a part of its energy to the recoil-proton by an elastic scattering.
It is referred to as the prompt event.
Then, the scattered neutron loses its energy by subsequent multiple scatterings,
and after $O(10)\mu{\rm s}$ it is captured by 
a gadolinium (Gd) nucleus in the Gd coated Mylar films wrapped around the scintillator.
The neutron capture on Gd results in a gamma cascade emission with
total energy of 7.9\,MeV for ${}^{157}{\rm Gd}$ and 8.5\,MeV for ${}^{155}{\rm Gd}$.
It is referred to as the delayed event.

We selected two kinds of delayed coincidence events,
correlated events with the delay time window of $8-150 \mu{\rm s}$ 
and accidental events with the time window of $1008-1150 \mu{\rm s}$.
The total energy is required to be $1.5\,{\rm MeV} \le E_{\rm total} \le 10.0\,{\rm MeV}$ 
for both the selections.
We calculated the number of neutron events by
subtracting the accidental event rate from the correlated event rate.

\begin{figure}[tbp]
  \begin{center}
    \includegraphics[width=9.5cm]{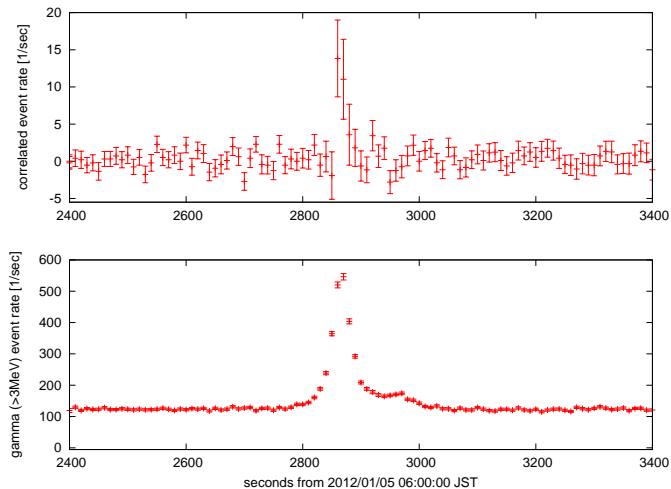}
    \caption{Net correlated event rate in the burst-20120105(upper panel) 
    			with the total event rate for the reference(bottom panel)}
    \label{fig:eventrates-neutron-burst20120105}
  \end{center}
\end{figure}

As a result, we found event-rate increase which is synchronized with
the $\gamma$-ray burst
with maximum rate of $14\pm 5 \,{\rm /s}$ in burst-20120105
(Figure~\ref{fig:eventrates-neutron-burst20120105}).

\section{Discussion}

Our observational results are mostly consistent with 
a model of the long duration $\gamma$-ray bursts from thunderclouds\cite{DwyerPR2014}.
In the model, seed electrons are provided to the thunderclouds
continuously mainly by secondary cosmic rays,
and they are multiplied by RREA process with a help of electric field between 
the negative charge of the lower part of a thundercloud 
and the positive pocket charge region just below the negative charge~\cite{Kitagawa1994}.
Then, those high energy electrons make elctromagnetic shower in the atmosphere resulting in numerous  $\gamma$-rays on the ground.
The downward vertical directions of $\gamma$-ray bursts also reflects the direction of the electric field.

The estimated almost monochromatic energies of the source electrons roughly reproduce the shape of the observed energy spectra.
However, the present analysis has little power to examine the theoretically favored model~\cite{Dwyer2011} in which the spectrum becomes exponential shape independent of the electric field or the air density.

The duration of the bursts may depend on
the movement or the development stage of the clouds.
For example, the duration of the burst-20120105, $\sim180\,{\rm s}$
can be explained by a thundercloud which have
typical velocity of 50\,km/h and diameter of 2.5\,km, 
which is somewhat smaller than typical echo size of thunderclouds 
(which is 4-6\,km~\cite{Kitagawa1994}).

In addition, due to the low temperature,
the altitude of the thunderclouds at midwinter
is low in the coastal area of the Japan Sea~\cite{Kitagawa1994},
which explains the result that 
the observed energy spectra implies the low altitude electron source.
Similar winter thunderclouds are relatively rare, but are also observed in the west coast of Norway
and toward the east coast from the Great Lakes of North America.

RREA and resulting electromagnetic shower may exist also in more common high altitude thunderclouds in summer.
However, $\gamma$-rays and electrons of shower may totally absorbed by the thick air under the clouds. 
It is also consistent with the observations that long duration bursts are reported in summer on high mountains,
where the relative altitude of the clouds are very low.

\section{Conclusion}

We observed three $\gamma$-ray bursts in winter
with the PANDA detector made of 360-kg plastic scintillator
at Ohi Power Station
which stands on the coastal area of the Japan Sea.
Table.~\ref{tab:features} summarizes the features of the observed bursts.
The maximum rate of the events with $E_{\rm total} \ge 3\,{\rm MeV}$
was $(5.5 \pm 0.1) \times 10^2 {\rm /s}$ 
of burst-20120105.

We found that 
for all the bursts periods,
there were active thunderclouds near the detector.

In addition,
we found that $\gamma$-rays of the bursts entered into the 
detector from the direction close to the zenith.
The arrival direction stayed constant during the burst.

These results indicated that the bursts originated in thunderclouds.
Monte Carlo simulation showed that
the observed $E_{\rm total}$ spectra of the bursts
are 
reproduced
by the bremsstrahlung $\gamma$-rays by
electrons with more or less monochromatic energy shown in
Figure~\ref{fig:electron_spectrum}
from low altitudes.

The arrival direction of the $\gamma$-rays and
the estimated energy of the source electrons of over $10\,{\rm MeV}$
can be described by relativistic runaway electron avalanche (RREA).
Namely, the secondary cosmic-ray electrons, which act as seed,
were accelerated and amplified in electric field of thunderclouds
by avalanche multiplication.

Additionally, neutrons were detected in one of the bursts
at the maximum rate of $\sim 14\pm5\, {\rm /s}$
with high confidence by the delayed coincidence method.
There is probability the event rate increase
includes neutrons emitted by the photonuclear reactions
on the nitrogen atoms in the air.
It could be due to the large flux of bremsstrahlung $\gamma$-ray 
with energy greater than the photonuclear reaction threshold for nitrogen.
The observation of fast neutrons on the ground implies
that more neutrons are produced in the air between a thundercloud and the ground
and even in the cloud itself.
However, only small fraction of those neutrons 
reach the ground because of the short absorption length
in the air\cite{Babich2013}. 

\begin{table*}[tbp]
  \begin{center}
    \caption{Features of the three $\gamma$-ray bursts}
    \label{tab:features}
    \begin{tabular}{p{6cm}|p{2.5cm}p{2.5cm}p{2.5cm}}\hline
      burst name& burst-20111225 & burst-20120102 & burst-20120105\\ \hline
      date of the burst & 
      2011, Dec. 25 & 2012, Jan. 2 & 2012, Jan. 5\\
      time(JST)& 
      05:07 & 09:19 & 06:46\\
      duration [s] &
      90 & 60 & 180 \\
      peak event rate ($E_{\rm tot} > 3{\rm MeV}$) [s$^{-1}$] & 
      $(2.3\pm0.1)\times 10^2$ & $(5.1\pm0.1)\times 10^2$
      & $(5.5\pm0.1)\times 10^2$\\
      correlated(neutron) events&
      not-detected & not-detected & detected\\
      \hline
    \end{tabular}
  \end{center}
\end{table*}

\section*{Acknowledgements}
The authors thank the Kansai Electric Power Co., Inc. for its cooperation for our experiment on site of Ohi Power Station.
The authors wish to acknowledge useful discussions with Dr. T. Torii and Dr. H.Tsuchiya.
This research was partially supported by the Japanese Ministry of Education, Science, Sports and Culture, Grant-in-Aid for COE Research, Grant-in-Aid for Scientific Research (B), Grant-in-Aid for Scientific Research on Innovative Areas, and Grant-in-Aid for JSPS Fellows,
Advanced Leading Graduate Course for Photon Science (ALPS) at the University of Tokyo, and also by the Mitsubishi Foundation.






\bibliographystyle{model1a-num-names}
\bibliography{k_bib}

\begin{thebibliography}{39}
\expandafter\ifx\csname natexlab\endcsname\relax\def\natexlab#1{#1}\fi
\providecommand{\url}[1]{\texttt{#1}}
\providecommand{\href}[2]{#2}
\providecommand{\path}[1]{#1}
\providecommand{\DOIprefix}{doi:}
\providecommand{\ArXivprefix}{arXiv:}
\providecommand{\URLprefix}{URL: }
\providecommand{\Pubmedprefix}{pmid:}
\providecommand{\doi}[1]{\href{http://dx.doi.org/#1}{\path{#1}}}
\providecommand{\Pubmed}[1]{\href{pmid:#1}{\path{#1}}}
\providecommand{\bibinfo}[2]{#2}
\ifx\xfnm\relax \def\xfnm[#1]{\unskip,\space#1}\fi
\bibitem[{Wilson(1924)}]{Wilson1924}
\bibinfo{author}{C.~T.~R. Wilson}, \bibinfo{journal}{Proc. Phys. Soc. London}
  \bibinfo{volume}{37} (\bibinfo{year}{1924}) \bibinfo{pages}{32D--37D}.
  \DOIprefix\doi{10.1088/1478-7814/37/1/314}.
\bibitem[{Fishman et~al.(1994)}]{TGF}
\bibinfo{author}{G.~J. Fishman}, et~al., \bibinfo{journal}{Science}
  \bibinfo{volume}{264} (\bibinfo{year}{1994}) \bibinfo{pages}{1313}.
\bibitem[{Dwyer et~al.(2015)}]{Dwyer2015}
\bibinfo{author}{J.~R. Dwyer}, et~al., \bibinfo{journal}{J. Plasma Physics}
  \bibinfo{volume}{81} (\bibinfo{year}{2015}) \bibinfo{pages}{475810405}.
\bibitem[{Suszcynsky et~al.(1996)Suszcynsky, Roussel-Dupre, and
  Shaw}]{Suszcynsky1996}
\bibinfo{author}{D.~M. Suszcynsky}, \bibinfo{author}{R.~Roussel-Dupre},
  \bibinfo{author}{G.~Shaw}, \bibinfo{journal}{J. Geophys. Res.}
  \bibinfo{volume}{101} (\bibinfo{year}{1996}) \bibinfo{pages}{23505--23516}.
  \DOIprefix\doi{10.1029/96JD02134}.
\bibitem[{Brunetti et~al.(2000)Brunetti, Cecchini, Galli, Giovannini, and
  Pagliarin}]{Brunetti2000}
\bibinfo{author}{M.~Brunetti}, \bibinfo{author}{S.~Cecchini},
  \bibinfo{author}{M.~Galli}, \bibinfo{author}{G.~Giovannini},
  \bibinfo{author}{A.~Pagliarin}, \bibinfo{journal}{Geophys. Res. Lett.}
  \bibinfo{volume}{27} (\bibinfo{year}{2000}) \bibinfo{pages}{1599--1602}.
  \DOIprefix\doi{10.1029/2000GL003750}.
\bibitem[{Chubenko et~al.(2000)Chubenko, Antonova, Kryukov, Piskal, Ptitsyn,
  Shepetov, Vildanova, Zybin, and Gurevich}]{Chubenko2000}
\bibinfo{author}{A.~Chubenko}, \bibinfo{author}{V.~Antonova},
  \bibinfo{author}{S.~Kryukov}, \bibinfo{author}{V.~Piskal},
  \bibinfo{author}{M.~Ptitsyn}, \bibinfo{author}{A.~Shepetov},
  \bibinfo{author}{L.~Vildanova}, \bibinfo{author}{K.~Zybin},
  \bibinfo{author}{A.~Gurevich}, \bibinfo{journal}{Phys. Lett. A}
  \bibinfo{volume}{275} (\bibinfo{year}{2000}) \bibinfo{pages}{90--100}.
  \DOIprefix\doi{10.1016/S0375-9601(00)00502-8}.
\bibitem[{Chubenko et~al.(2003)}]{Chubenko2003}
\bibinfo{author}{A.~Chubenko}, et~al., \bibinfo{journal}{Phys. Lett. A}
  \bibinfo{volume}{309} (\bibinfo{year}{2003}) \bibinfo{pages}{90}.
\bibitem[{Alexeenko et~al.(2002)Alexeenko, Khaerdinov, Lidvansky, and
  Petkov}]{Alexeenko2002}
\bibinfo{author}{V.~Alexeenko}, \bibinfo{author}{N.~Khaerdinov},
  \bibinfo{author}{A.~Lidvansky}, \bibinfo{author}{V.~Petkov},
  \bibinfo{journal}{Phys. Lett. A} \bibinfo{volume}{301} (\bibinfo{year}{2002})
  \bibinfo{pages}{299--306}. \DOIprefix\doi{10.1016/S0375-9601(02)00981-7}.
\bibitem[{Torii et~al.(2009)Torii, Sugita, Tanabe, Kimura, Kamogawa, Yajima,
  and Yasuda}]{Torii2009}
\bibinfo{author}{T.~Torii}, \bibinfo{author}{T.~Sugita},
  \bibinfo{author}{S.~Tanabe}, \bibinfo{author}{Y.~Kimura},
  \bibinfo{author}{M.~Kamogawa}, \bibinfo{author}{K.~Yajima},
  \bibinfo{author}{H.~Yasuda}, \bibinfo{journal}{Geophys. Res. Lett.}
  \bibinfo{volume}{36} (\bibinfo{year}{2009}) \bibinfo{pages}{L13804}.
  \DOIprefix\doi{10.1029/2008GL037105}.
\bibitem[{Tsuchiya et~al.(2009)Tsuchiya, Enoto, Torii, Nakazawa, Yuasa, Torii,
  Fukuyama, Yamaguchi, Kato, Okano, Takita, and Makishima}]{Tsuchiya2009}
\bibinfo{author}{H.~Tsuchiya}, \bibinfo{author}{T.~Enoto},
  \bibinfo{author}{T.~Torii}, \bibinfo{author}{K.~Nakazawa},
  \bibinfo{author}{T.~Yuasa}, \bibinfo{author}{S.~Torii},
  \bibinfo{author}{T.~Fukuyama}, \bibinfo{author}{T.~Yamaguchi},
  \bibinfo{author}{H.~Kato}, \bibinfo{author}{M.~Okano},
  \bibinfo{author}{M.~Takita}, \bibinfo{author}{K.~Makishima},
  \bibinfo{journal}{Phys. Rev. Lett.} \bibinfo{volume}{102}
  (\bibinfo{year}{2009}) \bibinfo{pages}{255003}.
  \DOIprefix\doi{10.1103/PhysRevLett.102.255003}.
\bibitem[{Tsuchiya et~al.(2012)Tsuchiya, Hibino, Kawata, Hotta, Tateyama,
  Ohnishi, Takita, Chen, Huang, Miyasaka, Kondo, Takahashi, Shimoda, Yamada,
  Lu, Zhang, Yu, Tan, Nie, Munakata, Enoto, and Makishima}]{Tsuchiya2012}
\bibinfo{author}{H.~Tsuchiya}, \bibinfo{author}{K.~Hibino},
  \bibinfo{author}{K.~Kawata}, \bibinfo{author}{N.~Hotta},
  \bibinfo{author}{N.~Tateyama}, \bibinfo{author}{M.~Ohnishi},
  \bibinfo{author}{M.~Takita}, \bibinfo{author}{D.~Chen},
  \bibinfo{author}{J.~Huang}, \bibinfo{author}{M.~Miyasaka},
  \bibinfo{author}{I.~Kondo}, \bibinfo{author}{E.~Takahashi},
  \bibinfo{author}{S.~Shimoda}, \bibinfo{author}{Y.~Yamada},
  \bibinfo{author}{H.~Lu}, \bibinfo{author}{J.~L. Zhang},
  \bibinfo{author}{X.~X. Yu}, \bibinfo{author}{Y.~H. Tan},
  \bibinfo{author}{S.~M. Nie}, \bibinfo{author}{K.~Munakata},
  \bibinfo{author}{T.~Enoto}, \bibinfo{author}{K.~Makishima},
  \bibinfo{journal}{Phys. Rev. D} \bibinfo{volume}{85} (\bibinfo{year}{2012})
  \bibinfo{pages}{092006}. \DOIprefix\doi{10.1103/PhysRevD.85.092006}.
\bibitem[{Chilingarian et~al.(2010)Chilingarian, Daryan, Arakelyan,
  Hovhannisyan, Mailyan, Melkumyan, Hovsepyan, Chilingaryan, Reymers, and
  Vanyan}]{Chilingarian2010}
\bibinfo{author}{A.~Chilingarian}, \bibinfo{author}{A.~Daryan},
  \bibinfo{author}{K.~Arakelyan}, \bibinfo{author}{A.~Hovhannisyan},
  \bibinfo{author}{B.~Mailyan}, \bibinfo{author}{L.~Melkumyan},
  \bibinfo{author}{G.~Hovsepyan}, \bibinfo{author}{S.~Chilingaryan},
  \bibinfo{author}{A.~Reymers}, \bibinfo{author}{L.~Vanyan},
  \bibinfo{journal}{Phys. Rev. D} \bibinfo{volume}{82} (\bibinfo{year}{2010})
  \bibinfo{pages}{043009}. \DOIprefix\doi{10.1103/PhysRevD.82.043009}.
\bibitem[{Chilingarian et~al.(2012)Chilingarian, Mailyan, and
  Vanyan}]{Chilingarian2012}
\bibinfo{author}{A.~Chilingarian}, \bibinfo{author}{B.~Mailyan},
  \bibinfo{author}{L.~Vanyan}, \bibinfo{journal}{Atmos. Res.}
  \bibinfo{volume}{114-115} (\bibinfo{year}{2012}) \bibinfo{pages}{1--16}.
  \DOIprefix\doi{10.1016/j.atmosres.2012.05.008}.
\bibitem[{Yamazaki et~al.(2002)Yamazaki, Fujimaki, Ohtaka, and
  Tonouchi}]{Yamazaki2002}
\bibinfo{author}{K.~Yamazaki}, \bibinfo{author}{H.~Fujimaki},
  \bibinfo{author}{T.~Ohtaka}, \bibinfo{author}{S.~Tonouchi},
  \bibinfo{journal}{Annual report of Niigata Prefectural Institute of Public
  Health and Environment Science} \bibinfo{volume}{17} (\bibinfo{year}{2002})
  \bibinfo{pages}{94--98}.
\bibitem[{Torii et~al.(2002)Torii, Takeishi, and Hosono}]{Torii2002}
\bibinfo{author}{T.~Torii}, \bibinfo{author}{M.~Takeishi},
  \bibinfo{author}{T.~Hosono}, \bibinfo{journal}{J. Geophys. Res.}
  \bibinfo{volume}{107} (\bibinfo{year}{2002}) \bibinfo{pages}{4324}.
  \DOIprefix\doi{10.1029/2001JD000938}.
\bibitem[{Enoto et~al.(2007)Enoto, Tsuchiya, Yamada, Yuasa, Kawaharada,
  Kitaguchi, Kokubun, Kato, Okano, Nakamura, and Makishima}]{Enoto2007}
\bibinfo{author}{T.~Enoto}, \bibinfo{author}{H.~Tsuchiya},
  \bibinfo{author}{S.~Yamada}, \bibinfo{author}{T.~Yuasa},
  \bibinfo{author}{M.~Kawaharada}, \bibinfo{author}{T.~Kitaguchi},
  \bibinfo{author}{M.~Kokubun}, \bibinfo{author}{H.~Kato},
  \bibinfo{author}{M.~Okano}, \bibinfo{author}{S.~Nakamura},
  \bibinfo{author}{K.~Makishima}, in: \bibinfo{booktitle}{ICRC 2007 Proc. -
  Pre-Conference Ed.}, volume~\bibinfo{volume}{1}, pp. \bibinfo{pages}{1--4}.
\bibitem[{Tsuchiya et~al.(2007)Tsuchiya, Enoto, Yamada, Yuasa, Kawaharada,
  Kitaguchi, Kokubun, Kato, Okano, Nakamura, and Makishima}]{Tsuchiya2007}
\bibinfo{author}{H.~Tsuchiya}, \bibinfo{author}{T.~Enoto},
  \bibinfo{author}{S.~Yamada}, \bibinfo{author}{T.~Yuasa},
  \bibinfo{author}{M.~Kawaharada}, \bibinfo{author}{T.~Kitaguchi},
  \bibinfo{author}{M.~Kokubun}, \bibinfo{author}{H.~Kato},
  \bibinfo{author}{M.~Okano}, \bibinfo{author}{S.~Nakamura},
  \bibinfo{author}{K.~Makishima}, \bibinfo{journal}{Phys. Rev. Lett.}
  \bibinfo{volume}{99} (\bibinfo{year}{2007}) \bibinfo{pages}{165002}.
  \DOIprefix\doi{10.1103/PhysRevLett.99.165002}.
\bibitem[{Tsuchiya et~al.(2011)Tsuchiya, Enoto, Yamada, Yuasa, Nakazawa,
  Kitaguchi, Kawaharada, Kokubun, Kato, Okano, and Makishima}]{Tsuchiya2011}
\bibinfo{author}{H.~Tsuchiya}, \bibinfo{author}{T.~Enoto},
  \bibinfo{author}{S.~Yamada}, \bibinfo{author}{T.~Yuasa},
  \bibinfo{author}{K.~Nakazawa}, \bibinfo{author}{T.~Kitaguchi},
  \bibinfo{author}{M.~Kawaharada}, \bibinfo{author}{M.~Kokubun},
  \bibinfo{author}{H.~Kato}, \bibinfo{author}{M.~Okano},
  \bibinfo{author}{K.~Makishima}, \bibinfo{journal}{J. Geophys. Res.}
  \bibinfo{volume}{116} (\bibinfo{year}{2011}) \bibinfo{pages}{D09113}.
  \DOIprefix\doi{10.1029/2010JD015161}.
\bibitem[{Tsuchiya et~al.(2013)Tsuchiya, Enoto, Iwata, Yamada, Yuasa,
  Kitaguchi, Kawaharada, Nakazawa, Kokubun, Kato, Okano, Tamagawa, and
  Makishima}]{Tsuchiya2013}
\bibinfo{author}{H.~Tsuchiya}, \bibinfo{author}{T.~Enoto},
  \bibinfo{author}{K.~Iwata}, \bibinfo{author}{S.~Yamada},
  \bibinfo{author}{T.~Yuasa}, \bibinfo{author}{T.~Kitaguchi},
  \bibinfo{author}{M.~Kawaharada}, \bibinfo{author}{K.~Nakazawa},
  \bibinfo{author}{M.~Kokubun}, \bibinfo{author}{H.~Kato},
  \bibinfo{author}{M.~Okano}, \bibinfo{author}{T.~Tamagawa},
  \bibinfo{author}{K.~Makishima}, \bibinfo{journal}{Phys. Rev. Lett.}
  \bibinfo{volume}{111} (\bibinfo{year}{2013}) \bibinfo{pages}{015001}.
  \DOIprefix\doi{10.1103/PhysRevLett.111.015001}.
\bibitem[{Torii et~al.(2011)Torii, Sugita, Kamogawa, Watanabe, and
  Kusunoki}]{Torii2011}
\bibinfo{author}{T.~Torii}, \bibinfo{author}{T.~Sugita},
  \bibinfo{author}{M.~Kamogawa}, \bibinfo{author}{Y.~Watanabe},
  \bibinfo{author}{K.~Kusunoki}, \bibinfo{journal}{Geophys. Res. Lett.}
  \bibinfo{volume}{38} (\bibinfo{year}{2011}) \bibinfo{pages}{L24801}.
  \DOIprefix\doi{10.1029/2011GL049731}.
\bibitem[{Gurevich et~al.(1992)Gurevich, Milikh, and
  Roussel-Dupre}]{Gurevich1992}
\bibinfo{author}{A.~Gurevich}, \bibinfo{author}{G.~Milikh},
  \bibinfo{author}{R.~Roussel-Dupre}, \bibinfo{journal}{Phys. Lett. A}
  \bibinfo{volume}{165} (\bibinfo{year}{1992}) \bibinfo{pages}{463--468}.
  \DOIprefix\doi{10.1016/0375-9601(92)90348-P}.
\bibitem[{Gurevich and Zybin(2005)}]{Gurevich2005}
\bibinfo{author}{A.~V. Gurevich}, \bibinfo{author}{K.~P. Zybin},
  \bibinfo{journal}{Phys. Today} \bibinfo{volume}{58} (\bibinfo{year}{2005})
  \bibinfo{pages}{37}. \DOIprefix\doi{10.1063/1.1995746}.
\bibitem[{Torii et~al.(2004)Torii, Nishijima, Kawasaki, and
  Sugita}]{Torii2004a}
\bibinfo{author}{T.~Torii}, \bibinfo{author}{T.~Nishijima},
  \bibinfo{author}{Z.-I. Kawasaki}, \bibinfo{author}{T.~Sugita},
  \bibinfo{journal}{Geophys. Res. Lett.} \bibinfo{volume}{31}
  (\bibinfo{year}{2004}) \bibinfo{pages}{L05113}.
  \DOIprefix\doi{10.1029/2003GL019067}.
\bibitem[{Babich et~al.(2010)Babich, Bochkov, Donskoi, and
  Kutsyk}]{Babich2010a}
\bibinfo{author}{L.~P. Babich}, \bibinfo{author}{E.~I. Bochkov},
  \bibinfo{author}{E.~N. Donskoi}, \bibinfo{author}{I.~M. Kutsyk},
  \bibinfo{journal}{J. Geophys. Res.} \bibinfo{volume}{115}
  (\bibinfo{year}{2010}) \bibinfo{pages}{A09317}.
  \DOIprefix\doi{10.1029/2009JA015017}.
\bibitem[{Dwyer and Uman(2014)}]{DwyerPR2014}
\bibinfo{author}{J.~R. Dwyer}, \bibinfo{author}{M.~A. Uman},
  \bibinfo{journal}{Phys. Rep.} \bibinfo{volume}{534} (\bibinfo{year}{2014})
  \bibinfo{pages}{147--241}.
\bibitem[{Shyam and Kaushik(1999)}]{Shyam1999}
\bibinfo{author}{A.~Shyam}, \bibinfo{author}{T.~C. Kaushik},
  \bibinfo{journal}{J. Geophys. Res.} \bibinfo{volume}{104}
  (\bibinfo{year}{1999}) \bibinfo{pages}{6867}.
  \DOIprefix\doi{10.1029/98JA02683}.
\bibitem[{Martin and Alves(2010)}]{Martin2010}
\bibinfo{author}{I.~M. Martin}, \bibinfo{author}{M.~A. Alves},
  \bibinfo{journal}{J. Geophys. Res.} \bibinfo{volume}{115}
  (\bibinfo{year}{2010}) \bibinfo{pages}{A00E11}.
  \DOIprefix\doi{10.1029/2009JA014498}.
\bibitem[{Gurevich et~al.(2012)Gurevich, Antonova, Chubenko, Karashtin, Mitko,
  Ptitsyn, Ryabov, Shepetov, Shlyugaev, Vildanova, and Zybin}]{Gurevich2012}
\bibinfo{author}{A.~V. Gurevich}, \bibinfo{author}{V.~P. Antonova},
  \bibinfo{author}{A.~P. Chubenko}, \bibinfo{author}{A.~N. Karashtin},
  \bibinfo{author}{G.~G. Mitko}, \bibinfo{author}{M.~O. Ptitsyn},
  \bibinfo{author}{V.~A. Ryabov}, \bibinfo{author}{A.~L. Shepetov},
  \bibinfo{author}{Y.~V. Shlyugaev}, \bibinfo{author}{L.~I. Vildanova},
  \bibinfo{author}{K.~P. Zybin}, \bibinfo{journal}{Phys. Rev. Lett.}
  \bibinfo{volume}{108} (\bibinfo{year}{2012}) \bibinfo{pages}{125001}.
  \DOIprefix\doi{10.1103/PhysRevLett.108.125001}.
\bibitem[{Starodubtsev et~al.(2012)Starodubtsev, Kozlov, Toropov, Mullayarov,
  Grigor'ev, and Moiseev}]{Starodubtsev2012}
\bibinfo{author}{S.~A. Starodubtsev}, \bibinfo{author}{V.~I. Kozlov},
  \bibinfo{author}{A.~A. Toropov}, \bibinfo{author}{V.~A. Mullayarov},
  \bibinfo{author}{V.~G. Grigor'ev}, \bibinfo{author}{A.~V. Moiseev},
  \bibinfo{journal}{JETP Lett.} \bibinfo{volume}{96} (\bibinfo{year}{2012})
  \bibinfo{pages}{188--191}. \DOIprefix\doi{10.1134/S0021364012150106}.
\bibitem[{Babich et~al.(2010)Babich, Bochkov, Kutsyk, and
  Roussel-Dupr\'{e}}]{Babich2010}
\bibinfo{author}{L.~P. Babich}, \bibinfo{author}{E.~I. Bochkov},
  \bibinfo{author}{I.~M. Kutsyk}, \bibinfo{author}{R.~A. Roussel-Dupr\'{e}},
  \bibinfo{journal}{J. Geophys. Res.} \bibinfo{volume}{115}
  (\bibinfo{year}{2010}) \bibinfo{pages}{A00E28}.
  \DOIprefix\doi{10.1029/2009JA014750}.
\bibitem[{Libby and Lukens(1973)}]{Libby1973}
\bibinfo{author}{L.~M. Libby}, \bibinfo{author}{H.~R. Lukens},
  \bibinfo{journal}{J. Geophys. Res.} \bibinfo{volume}{78}
  (\bibinfo{year}{1973}) \bibinfo{pages}{5902--5903}.
  \DOIprefix\doi{10.1029/JB078i026p05902}.
\bibitem[{Fleischer et~al.(1974)Fleischer, Plumer, and Crouch}]{Fleischer1974}
\bibinfo{author}{R.~L. Fleischer}, \bibinfo{author}{J.~A. Plumer},
  \bibinfo{author}{K.~Crouch}, \bibinfo{journal}{J. Geophys. Res.}
  \bibinfo{volume}{79} (\bibinfo{year}{1974}) \bibinfo{pages}{5013--5017}.
  \DOIprefix\doi{10.1029/JC079i033p05013}.
\bibitem[{Kuroda et~al.(2012)Kuroda, Oguri, Kato, Nakata, Inoue, Ito, and
  Minowa}]{LesserPANDA}
\bibinfo{author}{Y.~Kuroda}, \bibinfo{author}{S.~Oguri},
  \bibinfo{author}{Y.~Kato}, \bibinfo{author}{R.~Nakata},
  \bibinfo{author}{Y.~Inoue}, \bibinfo{author}{C.~Ito},
  \bibinfo{author}{M.~Minowa}, \bibinfo{journal}{Nuclear Instruments and Ms in
  Physics Reseach Section A} \bibinfo{volume}{690} (\bibinfo{year}{2012})
  \bibinfo{pages}{41--47}.
\bibitem[{Oguri et~al.(2014)Oguri, Kuroda, Kato, Nakata, Inoue, Ito, and
  Minowa}]{PANDA36}
\bibinfo{author}{S.~Oguri}, \bibinfo{author}{Y.~Kuroda},
  \bibinfo{author}{Y.~Kato}, \bibinfo{author}{R.~Nakata},
  \bibinfo{author}{Y.~Inoue}, \bibinfo{author}{C.~Ito},
  \bibinfo{author}{M.~Minowa}, \bibinfo{journal}{Nuclear Instruments and
  Methods in Physics Reseach Section A} \bibinfo{volume}{757}
  (\bibinfo{year}{2014}) \bibinfo{pages}{33--39}.
\bibitem[{{Japan Meteorological Agency}(????)}]{JapanMeteorologicalAgency}
\bibinfo{author}{{Japan Meteorological Agency}}, \bibinfo{title}{About kaminari
  nowcast - http://www.jma.go.jp/jma/kishou/know/toppuu/thunder2-1.html}.
\bibitem[{Agostinelli et~al.(2003)}]{Agostinelli2003}
\bibinfo{author}{S.~Agostinelli}, et~al., \bibinfo{journal}{Nucl. Instruments
  Methods Phys. Res. Sect. A Accel. Spectrometers, Detect. Assoc. Equip.}
  \bibinfo{volume}{506} (\bibinfo{year}{2003}) \bibinfo{pages}{250--303}.
  \DOIprefix\doi{10.1016/S0168-9002(03)01368-8}.
\bibitem[{Dwyer and Babich(2011)}]{Dwyer2011}
\bibinfo{author}{J.~R. Dwyer}, \bibinfo{author}{L.~P. Babich},
  \bibinfo{journal}{J. Geophys. Res.} \bibinfo{volume}{116}
  (\bibinfo{year}{2011}) \bibinfo{pages}{A09301}.
\bibitem[{Kitagawa and Michimoto(1994)}]{Kitagawa1994}
\bibinfo{author}{N.~Kitagawa}, \bibinfo{author}{K.~Michimoto},
  \bibinfo{journal}{J. Geophys. Res.} \bibinfo{volume}{99}
  (\bibinfo{year}{1994}) \bibinfo{pages}{10713--10721}.
  \DOIprefix\doi{10.1029/94JD00288}.
\bibitem[{Babich et~al.(2013)Babich, Bochkov, Kutsyk, and
  Zalyalov}]{Babich2013}
\bibinfo{author}{L.~P. Babich}, \bibinfo{author}{E.~I. Bochkov},
  \bibinfo{author}{I.~M. Kutsyk}, \bibinfo{author}{A.~N. Zalyalov},
  \bibinfo{journal}{JETP Letters} \bibinfo{volume}{97} (\bibinfo{year}{2013})
  \bibinfo{pages}{291--296}.

\end{thebibliography}







\end{document}